\documentclass[12pt]{article}
\pdfoutput=1
\usepackage[hidelinks]{hyperref}
\usepackage{xcolor}
\hypersetup{
    colorlinks,
    linkcolor={red!50!black},
    citecolor={blue!50!black},
    urlcolor={blue!80!black}
}

\usepackage[sorting=none]{biblatex}
\usepackage{amsmath}
\usepackage{times}
\usepackage{graphicx}
\usepackage{color}
\usepackage{multirow}
\usepackage{rotating}
\usepackage{bbm}
\usepackage{latexsym}
\title{
Kernel Heterogeneity Improves Sparseness of Natural Images Representations
}

\author{
Hugo J. Ladret$^{\displaystyle 1, \displaystyle 2}$,
Christian Casanova$^{\displaystyle 2}$,
Laurent Udo Perrinet$^{\displaystyle 1}$ 
\\
$^{\displaystyle 1}$ Institut de Neurosciences de la Timone, \\ UMR 7289, CNRS and Aix-Marseille Université, \\ Marseille, 13005, France\\
$^{\displaystyle 2}$ School of Optometry, Université de Montréal, \\ Montréal, QC H3C 3J7, Canada
}

\begin{document}

% \SetWatermarkText{preprint} % Text to be printed across the page
% \SetWatermarkScale{.5} % Size of the watermark text

\maketitle

% For example, continuous values can be sparsely represented by the mean and variance of the input's distribution. When dealing with complex-structured data, like natural images, these simplified approximations often lead to suboptimal performance. In these situations, it's often better to choose a less sparse approach that focuses on representing select parts of the input, along with their variance.
% ----------------------------------------------------------
% Abstract
% ---------------------------------------------------------
\begin{abstract}
Both biological and artificial neural networks inherently balance their performance with their operational cost, which balances their computational abilities. Typically, an efficient neuromorphic neural network is one that learns representations that reduce the redundancies and dimensionality of its input. This is for instance achieved in sparse coding, and sparse representations derived from natural images yield representations that are heterogeneous, both in their sampling of input features and in the variance of those features. Here, we investigated the connection between natural images' structure, particularly oriented features, and their corresponding sparse codes. We showed that representations of input features scattered across multiple levels of variance substantially improve the sparseness and resilience of sparse codes, at the cost of reconstruction performance. This echoes the structure of the model's input, allowing to account for the heterogeneously aleatoric structures of natural images. We demonstrate that learning kernel from natural images produces heterogeneity by balancing between approximate and dense representations, which improves all reconstruction metrics. Using a parametrized control of the kernels' heterogeneity used by a convolutional sparse coding algorithm, we show that heterogeneity emphasizes sparseness, while homogeneity improves representation granularity. In a broader context, these encoding strategy can serve as inputs to deep convolutional neural networks. We prove that such variance-encoded sparse image datasets enhance computational efficiency, emphasizing the benefits of kernel heterogeneity to leverage naturalistic and variant input structures and possible applications to improve the throughput of neuromorphic hardware. 
\end{abstract}


% ----------------------------------------------------------
% Introduction
% ----------------------------------------------------------
\section{Introduction}\label{sec:intro} 

\begin{figure}[h]
\centering
\includegraphics[width=\linewidth]{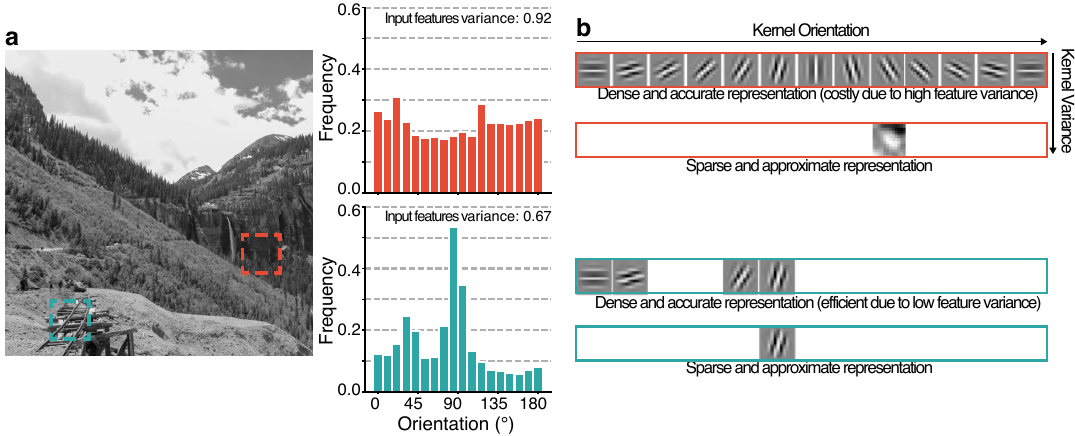}
\caption{
Efficient coding of sensory inputs.
\textbf{(a})~Orientation distributions with high (red) and low (blue) variance, in two $256^2$ pixel patches from a sample natural image.
\textbf{(b)}~Representation of these distributions and their efficiency depends on the structure of the input. 
The high-variance patch can be accurately represented with multiple oriented kernels, or approximated using one single kernel with high representational variance. Similarly, the low-variance patch can be encoded as a two-peaked orientation for an accurate representation, or using one kernel of low representation variance for a higher sparseness.
}
\label{fig:intro}
\end{figure}

% # General intro #
Neuromorphic neural networks are fundamentally designed to process inputs based on their statistical characteristics. This is particularly evident in vision-related tasks related to natural images, which exhibit a set of common statistical properties at multiple levels of complexity~\cite{simoncelli2001natural}. These statistical characteristics guide sensory processing, and are implicitly learned through efficient coding models~\cite{olshausen1996emergence, olshausen1997sparse}. For example, natural images typically show a local redundancy in luminance patterns that biological neural network remove at early processing stages, enhancing computational efficiency~\cite{laughlin1981simple}. In general, these images can be conceptualized as distributions of features (Figure~\ref{fig:intro}), which are, at a low descriptive level, oriented edges that form the foundation of hierarchical representations of natural images~\cite{boutin2020sparse}. The first moment of these distributions informs on the mean orientation in a given image patch, while the second central moment represents the heterogeneity of these features. 

% # Variance intro #
Modeling of such heterogeneity is crucial for sensory processing, both through input and representation bound variances~\cite{hullermeier2021aleatoric}. Input variance, also referred to as \textit{aleatoric} variance, stems from the intrinsic stochasticity in the processes that generate natural sensory inputs, such as sounds~\cite{nakamura2015robot}, textures~\cite{pettypiece2010integration} or images~\cite{ruderman1994statistics}. As its sources escape modeller control, it is challenging to predict, especially in computer vision models~\cite{gousseau2001natural} or neuromorphic hardware~\cite{fair2019neuromorphic}, and mandates a robust approach to accurately represent and process naturalistic inputs. 

% # Biological intro #
Evidences from neurobiological networks support the notion that neural systems account for this variance in decision-making processes~\cite{helmholtz1867treatise}, following Bayesian-derived rules~\cite{friston2005theory}. In practice, this is supported through the variability of neuronal sparse activations~\cite{orban2016neural}, which depends directly on the variance of the input~\cite{henaff2020representation,ladret2023cortical}. This relationship ties input variance to representational variance : in feature space, the basis function of a neuron is intrinsically linked to its capacity to encode particular levels of aleatoric variance~\cite{goris2015origin}. Neurons with broad kernels will more effectively encode broadly represented elements in orientation space, such as textures (see Figure~\ref{fig:intro}). This neurobiological evidence can notably serve to "explain away" irrelevant input to neural networks, thereby optimizing neuromorphic designs at the hardware level. 

% # ML intro #
Indeed, neuromorphic machine learning models which emulate the visual system, such as sparse coding, exhibit a dictionary of kernels which possess a wide range of tuning heterogeneity~\cite{olshausen1997sparse}. This heterogeneity is particularly notable in their convolutional forms, where feature activations, being both position- and scale-invariant, effectively mirror the aleatoric structure of natural images. This process is akin to maximum likelihood estimation, wherein modeling visual inputs involves capturing the variance of visual features through parametrized surrogate distributions.  Thus, sparse coding, with its minimalistic yet effective neuromorphic approximation of the early visual system, provides a valuable theoretical framework for understanding how input variance is tied to representational variance.

% # Intro Summary #
Here, we aim to provide an empirical account of this relationship, namely by showcasing the advantages of incorporating kernels with heterogeneous feature representations in sparse coding models of natural images. We use a convolutional sparse coding model, trained to reconstruct a novel dataset of high-definition natural images, and manipulate the heterogeneity of its kernels to study its reconstruction performances. We show that optimal learning relies on balancing the heterogeneity of features, which reflects the aleatoric variance in natural images. In a general context, we provide a full PyTorch implementation of our convolutional sparse coding algorithms, and use these codes as inputs of a deep convolutional network, boosting resilience to adversarial input degradation. This underscores our finding that inherent heterogeneity of kernels in machine learning, akin to that of receptive fields in biology,  enhances computational efficiency by effectively mirroring the statistical properties of inputs. % # research summary

% ----------------------------------------------------------
% Methods
% ----------------------------------------------------------
\section{Methods}\label{sec:methods} 

\subsection{Convolutional Sparse Coding}\label{sec:CSC}  
Sparse coding (SC) is an unsupervised method for learning the inverse representation of an input signal~\cite{lee2006efficient}. Given the assumption that a signal can be represented as a linear mixture of kernels (or basis functions), SC aims to minimize the activation of kernels used to represent the input signal, yielding an efficient representation~\cite{perrinet2015sparse} that can be inverted for reconstruction. Here, SC was used to reconstruct an image $s$ from sparse representations $x$, while minimizing the $\ell_1$-norm of the representation:
\begin{equation} \label{eq:sc}
    \underset{x}{\operatorname{arg min}} \frac{1}{2} || s - D x ||^2_2 + \lambda ||x ||_1
\end{equation} 
where $D$ is the set of kernels used to represent $s$ (called a dictionary) and $\lambda$ a regularization parameter that controls the trade-off between fidelity and sparsity. Conveniently, this problem can be efficiently approached with a Basis Pursuit DeNoising (BPDN) algorithm~\cite{chen2001atomic}. As there is \textit{a priori} no topology among elements of the dictionary, SC does not preserve the spatial structure of the input signal, which can be problematic in the context of the representation of natural images. Moreover, the overall decomposition is applied globally and handles poorly the overlap between redundant statistical properties of patches in the image~\cite{simoncelli2001natural}, yielding a suboptimal representation of the input signal~\cite{lewicki1998coding}.

These problems are leveraged by Convolutional sparse coding (CSC), an extension of the SC method to a convolutional representation, which is closer to a rough neurally-inspired design~\cite{serre2007feedforward} as used in deep convolutional network (CNNs)~\cite{boutin2022}. These CNNs use localized kernels similar to the receptive fields of biological neurons in the primary visual cortical areas. A convolutional architecture uses convolutional kernels (dictionary elements) that are spatially localized and replicated on the full input space (or possibly with a stride which subsamples that space). The number of kernels in the dictionary defines the number of features, or \textit{channels}. In CSC, the total number of kernels with respect to standard SC is multiplied by the number of positions. As a result, a convolution allows to explicitly represent the spatial structure of the signal to be reconstructed. This further reduces the number of kernels required to achieve an efficient representation of an image, while providing shift-invariant representations.  CSC extends equation (\ref{eq:sc}) to:
\begin{equation} \label{eq:csc}
    \underset{\{x_k\}}{\operatorname{arg min}} \frac{1}{2}
    || s - \sum_{k=1}^K \text{d}_k \ast x_k ||^2_2 + 
    \lambda \sum_{k=1}^K ||x_k||_1
\end{equation}
where $x_k$ is a $N^2$ dimensional coefficient map (given a $N^2$ sized image), $\text{d}_k$ is one kernel (among $K$ channels) and $\ast$ is the convolution operator. As the convolution is a linear operator, CSC problems can be solved with convolutional BPDN algorithms~\cite{wohlberg2015efficient}. Here, we used the Python SPORCO package~\cite{wohlberg2017sporco} to implement CSC methods, using an Alternating Direction Method of Multipliers (ADMM) algorithm~\cite{wang2019global} which splits Convolutional Sparse Coding problems into two alternating sub-problems, as described in Appendix A. Additionally, CSC proves advantageous over other reconstruction techniques in its ability to learn interpretable and visualizable kernels from input data.

% ----------------------------------------------------------

\subsection{Dictionaries}\label{sec:dicts} 
Optimal dictionaries to reconstruct natural images are known to be localized, oriented elements~\cite{hubel1962receptive, olshausen1996emergence}. Here, we utilized log-Gabor filters, which have been shown to accurately model the receptive fields of neurons in the visual cortex. These filters have several advantages compared to Gabor filters, notably that they do not have a DC component and that they optimally capture the log-frequency structure of natural images to ensure its optimal reconstruction~\cite{fischer2007sparse}. The log-Gabor filter~\cite{fischer2007self} is defined in the frequency domain by polar coordinates $(f, \theta)$  as:
\begin{equation} \label{eq:loggabor}
    G(f, \theta) = \exp \left( - \frac 1 2 \cdot
    \frac{\log(f/f_0)^2}{\log(1 + \sigma_f/f_0)^2}
    \right)
    \cdot
%    \exp \left( \frac{-(\theta - \theta_0)^2}{2\sigma^2_\theta} \right)
    \exp \left( \frac{\cos(2 \cdot (\theta - \theta_0))}{4 \cdot \sigma^2_\theta}  \right)
\end{equation}
where $f_0$ is the center frequency, $\sigma_f$ the bandwidth parameter for the frequency, $\theta_0$ the center orientation and $\sigma_\theta$ the standard deviation for the orientation. This provides with a parametrization of the dictionary, which is useful to compare the efficiency of different sparse coding models~\cite{fischer2006sparse}. We kept $f_0 = \sigma_f = 0.4$ cpd, varying only the orientation-related parameters to build the dictionaries. The angular bandwidth $B_\theta$ of the log-Gabor filter, expressed in degrees, was defined as $B_\theta = \sigma_\theta \sqrt{2\log2}$~\cite{swindale1998orientation}.

To titrate the impact of including heterogeneity in the dictionary, we created two log-Gabor dictionaries with the same number of  channels, one with homogeneous (a single $\sigma_\theta$) the other with heterogeneous (multiple $\sigma_\theta$) variance of representations. We compared these dictionaries before and after fine-tuning on the dataset, using a dictionary learned from scratch over the dataset as a fifth reference. Such learning was done by performing convolutional sparse coding in a multi-image setting:
\begin{equation} \label{eq:mcsc}
    \underset{\{\text{x}_k,j\}}{\operatorname{arg min}} \frac{1}{2}
    \sum_{j=1}^J
    || \sum_{k=1}^K \text{d}_k * \text{x}_{k,j} - s_j ||^2_2 + 
    \lambda \sum_{k}^{K} \sum_{j}^{J} ||\text{x}_{k,j}||_1
    \text{ s.t.\  $\forall k$, $||$d$_k||_2 = 1$}
\end{equation}
where $s_j$ is the $j$-th image in the dataset and $\text{x}_{k,j}$ is the coefficient map for the $k$-th filter and the $j$-th image. This was alternated with an optimization step of the dictionary: 
\begin{equation} \label{eq:learning}
\min_{D} \sum_{i=1}^N \frac{1}{2} \|x_i - D * z_i\|^2_2
\end{equation}
subject to the constraint $|d_k|_2 \le 1$ for $k = 1, \dots, K$. 

Performance of these dictionaries was measured with two metrics. The peak signal-to-noise ratio (PSNR), a common metric to evaluate reconstruction quality of grayscale images, is defined as:
\begin{equation}
    \text{PSNR}(I_1, I_2) = 20 \cdot \log_{10}(\text{max}(I_1))
    - 10 \cdot \log_{10} 
    \left( \frac{1}{m\cdot n} \sum_{i=1}^{m}\sum_{j=1}^{n} (I_1 - I_2)^2 \right)
\end{equation}
where $\text{max}(I_1)$ is the maximum pixel intensity of the source image. The right hand-side term of the PSNR is the $\log_{10}$ of the mean squared error, where $I_1$ and $I_2$ represent the pixel intensity in the source and reconstructed images, respectively. Given that the natural images used here are encoded on $8$ bits, common values of PSNR range between $20$ (worse) to $50$ (best) dB. We also measured the sparseness of the algorithm, which was defined as the fraction of basis coefficients used in a reconstruction which are equal to zero. This value is between $0$ (no nonzero coefficient) and $1$ (all coefficients are zero). Parametrization of the algorithm was chosen to balance sparseness and PSNR (Appendix A), i.e. $\lambda=0.05$, with $750$ iterations of the learning phase, a residual ratio of $1.05$ with relaxation at $1.8$, and dictionaries with $K=144$ total elements of $12^2$ pixels each.

% ----------------------------------------------------------

\subsection{Histogram of oriented gradients}\label{sec:hog}
The distributions of oriented features in Figure~\ref{fig:intro} were computed using a histogram of gradient orientations. Using the `scikit-image` library~\cite{van2014scikit}, given an input image \( I \) of dimension \( M \times N \), two gradients were computed at each pixel using Sobel filters $G_h(x,y)$ and $G_v(x,y)$, respectively, for vertical and horizontal gradients. The maps of the magnitude \( G_m \) and direction \( \theta \) were then given as:
\begin{equation}
    \begin{aligned}
        G_m(x,y) &= \sqrt{G_h(x,y)^2 + G_v(x,y)^2} \\
        \theta(x,y) &= \arctan2(G_v(x,y), G_h(x,y))
    \end{aligned}
\end{equation}
The range of possible gradient directions over \( [0, \pi] \) was divided into 18 bins. The orientation histogram \( H \) for each bin \( b \) was computed as:
\begin{equation}
H(b) = \sum_{(x,y)} I_b(\theta(x,y))
\end{equation}
where \( I_b \) is an indicator function, ranging from 1 if \( \theta(x,y) \) falls within the range of the bin \( b \) and $0$ otherwise. In that context, one can quantify the orientation content in natural images, then estimate the distribution of oriented features within the input: aleatoric variance can then be approximated as the inverse of the squared variance of this distribution in orientation space and is computed as $\text{Var}_{\text{circ}} = 1 - \sqrt{\bar{X}^2 + \bar{Y}^2}$, where \( \bar{X} \) and \( \bar{Y} \) are the average cosine and sine values respectively, yielding a scalar value between $0$ (lowest orientation variance) and $1$ (highest). 

% ----------------------------------------------------------

\subsection{Dataset}
Images for the CSC sections were captured using either a Canon EOS 650D or Canon EOS 6D camera, fitted with 28mm lenses. A total of $1145$ images was collected at a resolution of at least $5184\times3456$ pixels. For CSC, we extracted and used the central $256\times256$ pixel segment of each image. These images represent a variety of dynamic scenarios, and were carefully shot to ensure that the subjects of interest were in focus and entirely within the frame. We have made this dataset publicly available on Figshare~\cite{ladret2023imgs}.

% ----------------------------------------------------------

\subsection{Image classification using deep learning}\label{sec:dataset}
To evaluate the role of sparse codes obtained, we decided to go further than only measuring representation performance by applying these codes on a common machine learning task: image classification. To perform such classification in a neuromorphic-inspired setting, we utilized a modified version of the CIFAR-10 dataset. This dataset, which is commonly used for image classification, originally contains $60,000$ color images of $32\times32$ pixel resolution across $10$ balanced classes. We processed these images by first upscaling them to $128\times128$ resolution via bilinear interpolation. Subsequently, they were converted to grayscale and sparse-coded, as described above. 

The dataset was divided into a training set containing $50,000$ sparse codes and a test set comprising $10,000$ sparse codes. The network was trained from scratch through a standard PyTorch implementation, with backpropagation of the gradient using the Adam optimizer~\cite{kingma2014adam}. The training objective was to minimize the categorical cross-entropy loss, defined as:
\begin{equation}
    J(\theta) = -\frac{1}{N} \sum_{i=1}^{N} \sum_{j=1}^{C} y_{ij} \log(\hat{y}_{ij})
\end{equation}
where $N$ is the number of samples, $C$ is the number of classes, $y_{ij} $ is the true label, and $ \hat{y}_{ij} $ is the predicted label. 
The Adam update rule for each parameter \( \theta \) is based on moment estimates given by:
\begin{equation}
\theta_{t+1} = \theta_t - \eta \cdot \frac{\hat{m}_t}{\sqrt{\hat{v}_t} + \epsilon}
\end{equation}
where \( \eta \) is the learning rate, \( \hat{m}_t \) and \( \hat{v}_t \) are estimates of the mean and variance of the gradients, and \( \epsilon \) is a small constant to prevent division by zero.

The sparse codes representing these images were then used as inputs for an adapted ResNet-18 architecture~\cite{he2016deep} which is a classically used CNN architecture. This deep residual neural network, typically composed of 18 layers and used for various vision tasks, was adapted to process the $144$ dimensions of the sparse-coded inputs instead of the standard 3-channel (RGB) format. This dimensionality corresponds to the number of channels in our sparse coding dictionary. No other modifications were implemented in the network architecture design.

Hyperparameters were tuned via grid search to maximize accuracy on heterogeneous variance codes, with the resulting values: $\eta = 2e-4$, $\hat{m}_t = 0.9$, $\hat{v}_t = 0.99$, $\epsilon=1e-08$. When training the network, CSC methods using ADMM algorithms were ported from SPORCO to a custom PyTorch implementation (available at 
\href{https://github.com/hugoladret/epistemic_CSC}{\url{https:/github.com/hugoladret/epistemic_CSC}}) 
% \href{XXX}{XXX}) 
to speed up computations. 

% ----------------------------------------------------------
% Results
% ----------------------------------------------------------
\section{Results} 

% ----------------------------------------------------------
\subsection{Heterogeneous kernels improve the sparseness of natural images representations}
\begin{figure*}[h]
\centering
\includegraphics[width=\linewidth]{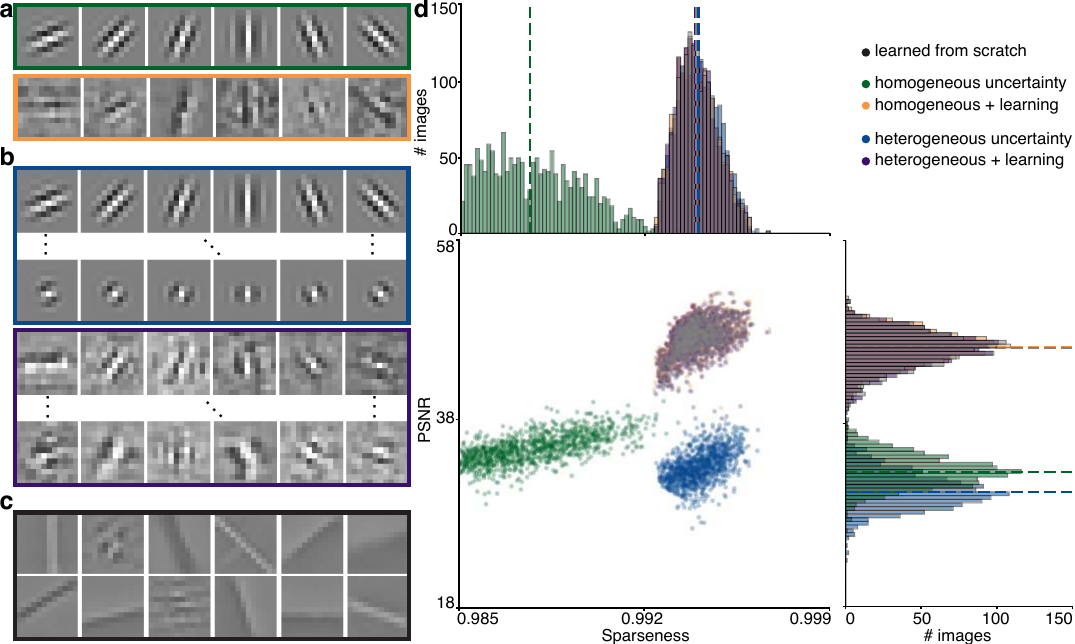}
\caption{
Kernel heterogeneity and reconstruction trade-off.
\textbf{(a)} Elements from dictionaries with homogeneous kernel variance before (green) and after dictionary learning (orange).
\textbf{(b)} Same, with heterogeneous kernel variance before (blue) and after learning (purple).
\textbf{(c)} Elements from a dictionary learned from random initialization on the dataset.
\textbf{(d)} Distribution of the sparseness (top) and Peak Signal-to-Noise Ratio (PSNR, right) of the five dictionaries. Median values are shown as dashed lines. All three post-learning dictionaries have overlapping (but not identical) distributions.
}
\label{fig:dicos}
\end{figure*}

We explored how variance in sensory inputs and neuromorphic representations controls the encoding strategies of natural images.
We compared five distinct convolutional sparse coding dictionaries of similar sizes. Two dictionaries using Log-Gabor filters were constructed : one with a homogeneous level of orientation variance ($B_\theta=12.0$°) and 72 orientations $\theta_0$ ranging from $0$° to $180$° (Figure~\ref{fig:dicos}a, green) compared to another one with heterogeneous orientation variance, spanning 12 orientation values $\theta_0$ and six $B_\theta$ ranging from $3$° to $30$° (Figure~\ref{fig:dicos}b, blue).
We then benchmarked these constructed dictionaries against their learned counterparts, which were fine-tuned on the dataset (Figure~\ref{fig:dicos}a, orange; b, purple). A final comparison was made against a randomly initialized dictionary learned \textit{de novo} on the same dataset (Figure~\ref{fig:dicos}c, black).
Performance evaluation across the $1,445$ high-definition natural images revealed that dictionaries initialized with Log-Gabor filters consistently displayed highly variant performance from image to image (Figure~\ref{fig:dicos}d). Prior to learning, the dictionary integrating heterogeneous orientation variance outperformed its homogeneous counterpart in sparsity (Mann-Whitney U-test, $U=1310760.0$, $p<0.001$), but had significantly lower PSNR ($U=262261.0$, $p<0.001$). Post-learning, all dictionaries had similar performances in terms of both sparsity ($U=634605$, $p=0.18$ for homogeneous vs random initialized dictionaries ; $U=634605.0$, $p=0.97$ for heterogeneous vs random initialized dictionaries) and PSNR ($U=694175$, $p=0.46$ ; $U=653943.0$, $p=0.99$). This suggests that emphasis on heterogeneous variance modelling improves the sparsity, at the cost of reconstruction performance. 

After learning from the dataset, whether from random initialization or from a pre-constructed log-Gabor dictionary, all dictionaries converge to qualitatively quite different filters, yet with a similar, superiorly sparse and performant form of encoding. The learning method indeed enhanced all Log-Gabor dictionaries, resulting in increased PSNR ($U=0.0$, $p<0.001$ ; $U=181535.0$, $p<0.001$, homogeneous and heterogeneous variance dictionaries, compared to their pre-learning version) and sparseness ($U=23595.0$, $p<0.001$ ; $U=248667.0$, $p<0.001$). Given the converging reconstruction and sparseness for all these dictionaries, we now focus on the heterogeneous variance dictionary, both pre- and post-learning, as well as the pre-learned homogeneous variance dictionary. Additional performance details for the homogeneous dictionary are provided in Appendix B.

\begin{figure}[t!]
\centering
\includegraphics[width=\linewidth]{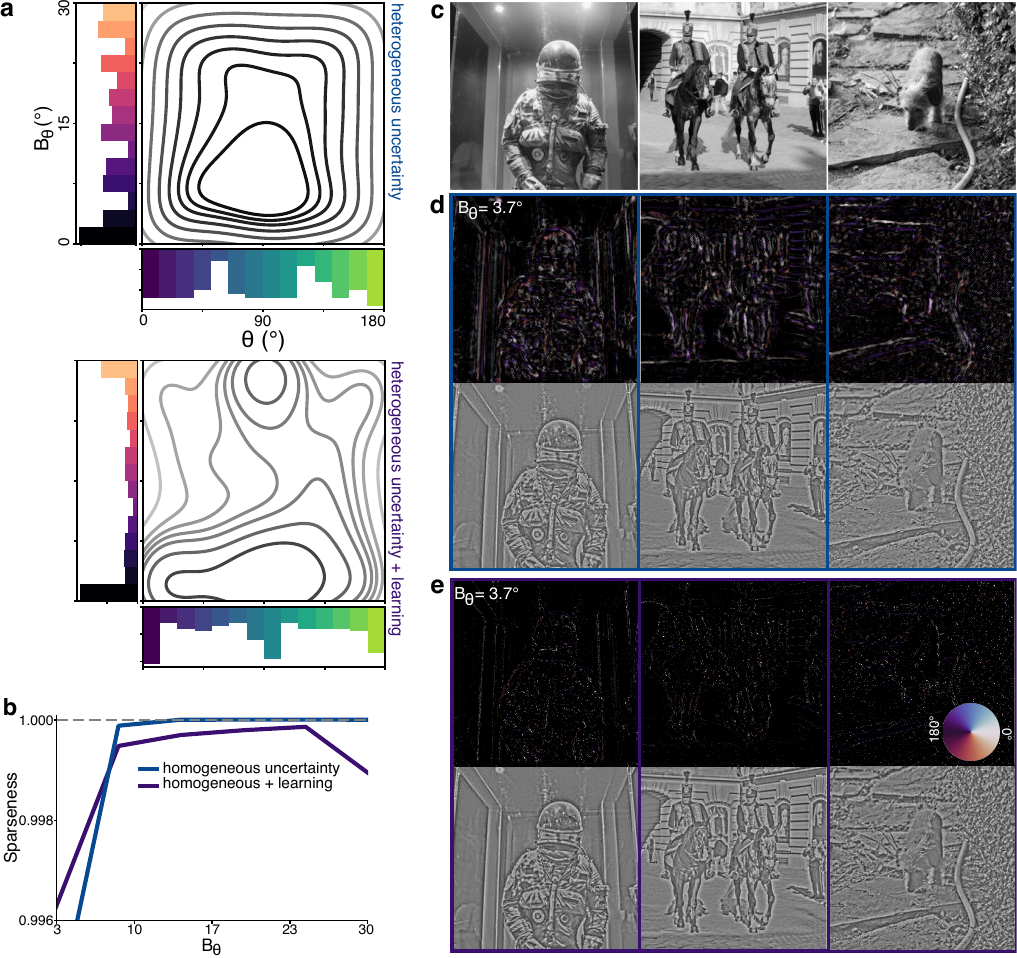}
\caption{
Learning balances coefficient distribution.
\textbf{(a)} Kernel density estimation over $\theta$ and $B_\theta$ of the kernels before (top) and after (bottom) learning.
\textbf{(b)} Sparseness of the dictionaries for kernel variance $B_\theta$. Sparseness $=1$ (i.e. no activation, as in the case of the pre-learning encoding) is represented as a gray dashed line.
\textbf{(c)} Example images from the dataset.
\textbf{(d)} Sparse code for high $B_\theta$ values (color coded by each coefficient's $\theta$) and reconstructions for the pre-learned, heterogeneous variance dictionary. 
\textbf{(e)} Same as (d), for post-learned, heterogeneous variance dictionary. Orientation color code of the coefficients is shown on the rightmost coefficient map.
}
\label{fig:learning}
\end{figure}

What are then the kernel features changed through the learning process? While fine-tuned dictionaries do incur a significantly higher computational cost during the learning phase, they deliver substantial improvements in both PSNR and sparsity, compared to merely introducing heterogeneous variance into a pre-existing dictionary. These enhancements can be attributed to modifications in the dictionary coefficients following the learning phase, affecting both the feature orientations ($\theta_0$) and their associated levels of variance ($B_\theta$) (Figure~\ref{fig:dicos}a).
Specifically, learning from a dataset of natural images introduced a bias toward cardinal orientations (Figure~\ref{fig:learning}a), mirroring inherent biases found in natural scenes~\cite{appelle1972perception}, which is in contrast to the uniformly distributed initial dictionary. Furthermore, the learning process resulted in a non-uniform distribution of coefficients across multiple levels of orientation variance (Figure~\ref{fig:learning}b). Notably, coefficients that were previously inactive (i.e., sparseness $=1$) became activated at higher $B_\theta$ levels (Figure~\ref{fig:learning}c-e). This led to consistent patterns in coefficient distribution across heterogeneous variance levels (Figure~\ref{fig:learning}d,e).
This uniformity is likely influenced by the dataset's inherent variability. Consequently, the performance gains attributed to the learning process are contingent upon feature orientation biases ($\theta_0$) and a redistribution of the levels of variance ($B_\theta$), both of which should be reflective of the dataset's intrinsic structure.

% ----------------------------------------------------------
\subsection{Statistical properties of natural images reflect the variance of learned sparse code}
\begin{figure*}[ht!]
\centering
\includegraphics[width=\linewidth]{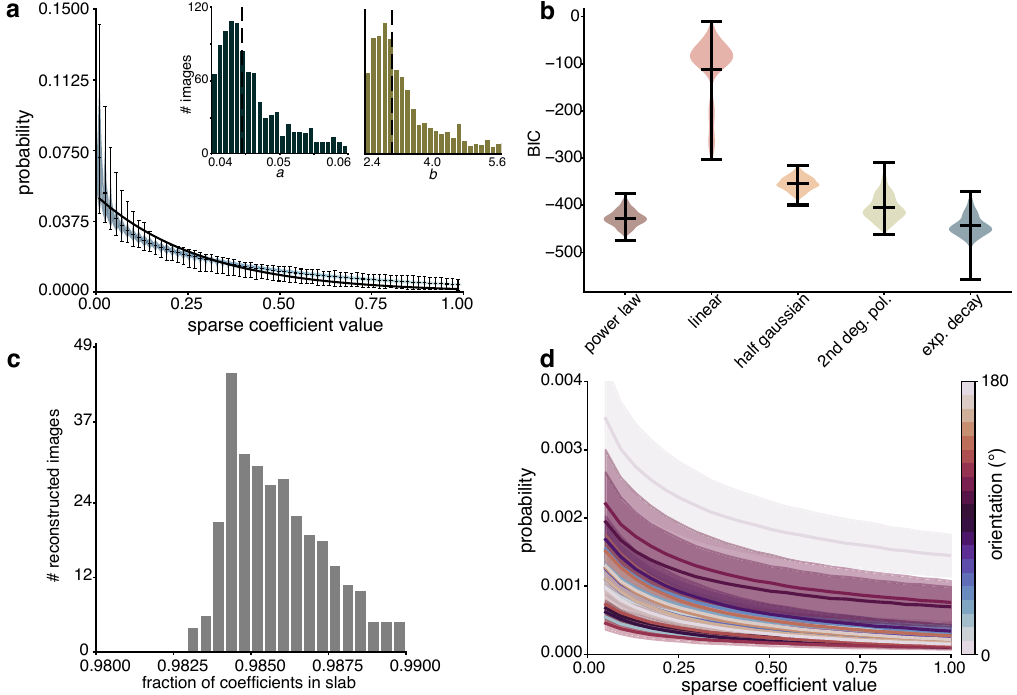}
\caption{
Spike-and-slab sparse representation of the natural images.
\textbf{(a)} Distribution of the sparse coefficients values. Violin plots' central lines represent mean values, with top and bottom lines representing the extrema. For each image, this distribution was fitted with an exponential decay (black line) $y = a\cdot\exp(-b\cdot x)$, with the distributions for the parameters over the $1145$ images shown in inset
\textbf{(b)} Bayesian Information Criterion (BIC) for the fitting of the distribution of spikes coefficients with different alternative functions. 
\textbf{(c)} Proportion of zero coefficients per image, i.e., belonging to the "spike" of the distribution.
\textbf{(d)} Same as (a), with coefficients split by different encoded orientation.
}
\label{fig:slab}
\end{figure*}

The criteria for the relevance of features encoded in neural networks is dictated by the statistical properties of the environment itself~\cite{ruderman1994statistics,simoncelli2001natural}. For instance, at a fundamental representational level, the neural code for light patterns in the retina is the cumulative sum of the Gaussian distribution of luminance found in natural images~\cite{laughlin1981simple}. At higher levels, scale distributions of visual features, in the Fourier domain, obey a $1/f^2$ power law, which once again echoes the power-law behavior of cortical responses~\cite{field1987relations, stringer2019high}. At intermediate levels, the distribution of these oriented edges can be characterized along its first- and second-order moments: a median orientation, and its corresponding variance. A proper model of natural images thus depends on a proper model of both these moments, which is reflected in the response properties of primary visual cortex neurons~\cite{hubel1962receptive}. Which of these two parameters warrants greater emphasis? Previous studies suggested that heterogeneity on both orientation and variances arises from sparse learning processes, \textit{in silico}~\cite{olshausen1996emergence} and \textit{in vivo}~\cite{goris2015origin}.

\begin{figure*}[t!]
\centering
\includegraphics[width=\linewidth]{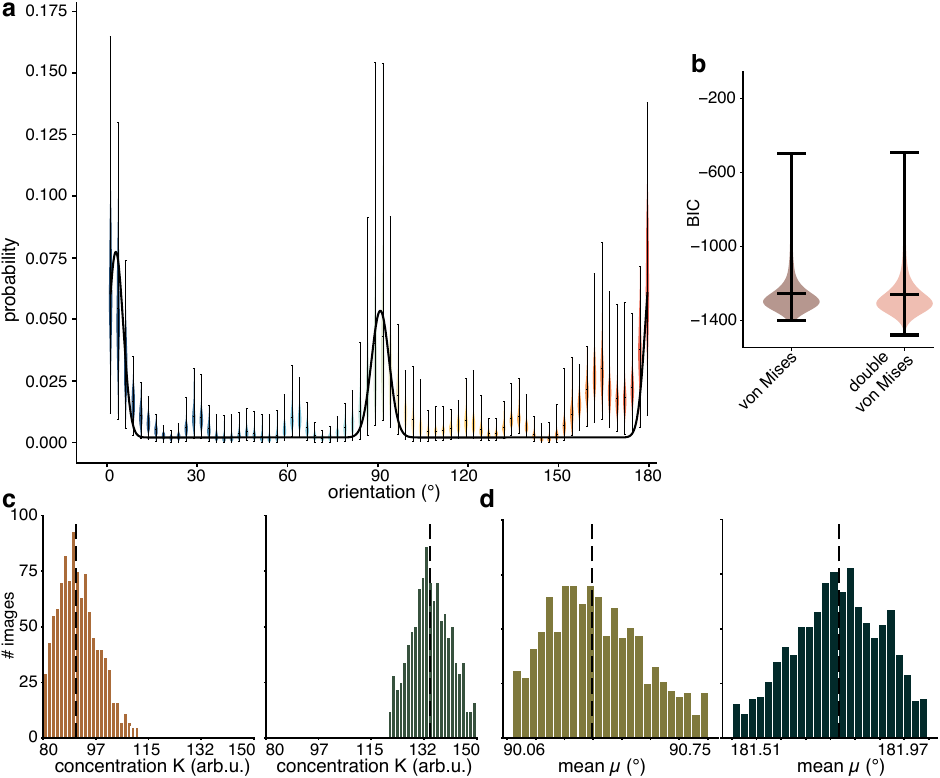}
\caption{
Orientations in natural images follow a double von Mises distribution.
\textbf{(a)} Orientations of the sparse coefficients, fitted with a double von Mises distribution (black line).
\textbf{(b)} Bayesian Information Criterion (BIC) for the fitting of the distribution of orientation coefficients.
\textbf{(c)} Distribution of the concentration parameter $\kappa$ for the first (left) and second (right) peaks of the double von Mises distribution.
\textbf{(d)} Same as (c), for the mean parameter $\mu$.
}
\label{fig:orientations}
\end{figure*}

Inherently, sparse coding enforces a prior on using a minimal number of coefficients to reconstruct an image, and is thus an encoding strategy that produces a "spike and slab" distribution of activations, characterized by a predominance of zero coefficients~\cite{field1987relations} (Figure~\ref{fig:slab}a-c). This imposes a prior on the representation of images at the feature-level, with a decaying exponential variation of coefficients that unfolds heterogeneously across different types of orientations (Figure~\ref{fig:slab}d). Lower BIC indicate less information lost in the fitting process, and thus a better fit.
Such heterogeneity in feature space stems from the fact that orientations in natural images are biased to cardinal (i.e., vertical and horizontal) orientations~\cite{coppola1998distribution}, which is echoed at the neuronal level by a cardinal bias in visual perception~\cite{hansen2004horizontal}. This biased distribution of orientation is well-captured by a double von Mises distribution in orientation space (Figure~\ref{fig:orientations}a,b): 
\begin{equation}
        f(x) = A_1 \exp\left( k_1 \left( \cos\left( 2 \pi  (x - \phi_1) \right) - 1 \right) \right)
        + A_2 \exp\left( k_2 \left( \cos\left( 2 \pi (x - \phi_2) \right) - 1 \right) \right)
\end{equation}
where \( A_1, A_2 \) are the amplitudes of the two von Mises distributions, \( k_1, k_2 \) are the concentration parameters for the two distributions, \( \phi_1, \phi_2 \) are the phase offsets for the two distributions. 

This distribution is known for higher heterogeneity, and thus aleatoric variance, in natural images compared to synthetic ones~\cite{coppola1998distribution}. At the cardinal orientations, this is also captured by the variation of the concentration parameters (Figure~\ref{fig:orientations}c,d) of the von Mises distributions, which underlies the notion that a proper description of natural images must be able to account for heterogeneous levels of aleatoric variance. This mandates a comparative evaluation of performance between dictionaries that emphasize a representation based on homogeneous or heterogeneous strategies, that is, emphasizing encoding mean features or their variances.

% ----------------------------------------------------------
\subsection{Heterogeneity improves resilience of the neural code}
\begin{figure}[t!]
\centering
\includegraphics[width=\linewidth]{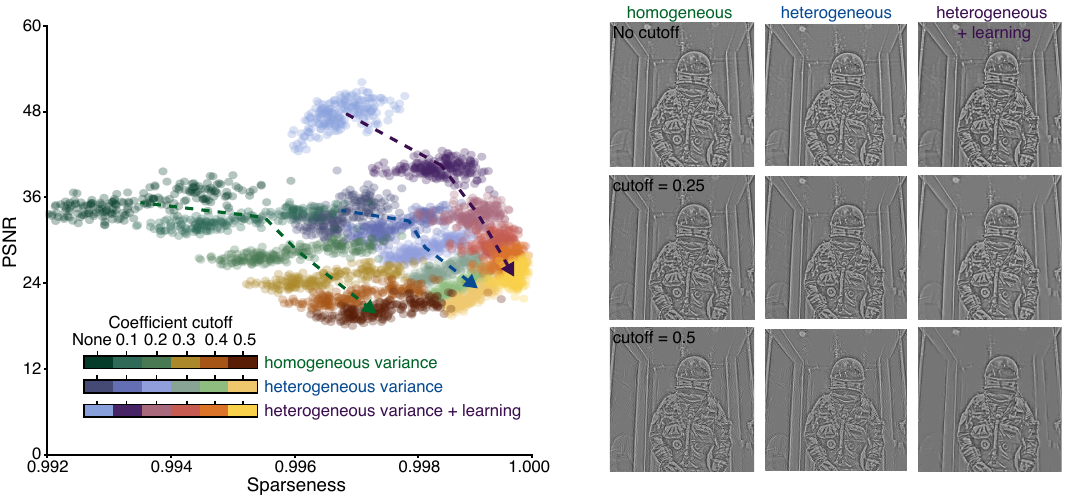}
\caption{
Sparse coefficients can be pruned for increased sparsity.
\textbf{(a)} Pruning of the coefficients based on their values and resulting sparseness/PSNR for three dictionaries, with mean trajectory represented as a dashed arrow.
\textbf{(b)} Reconstruction of an image with different cutoff levels.
}
\label{fig:cutoff}
\end{figure}

\begin{figure}[t!]
\centering
\includegraphics[width=\linewidth]{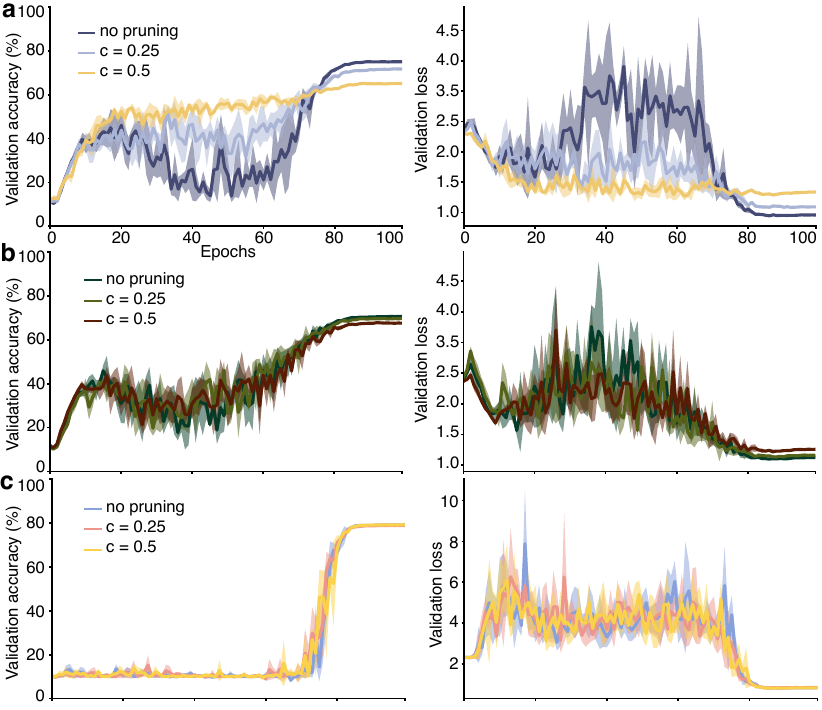}
\caption{
Deep Neural Networks (here, ResNet18), can be trained on sparse codes.
\textbf{(a)} Validation accuracy (left) and losses (right) curves, for $3$ different pruning levels of coefficients for the heterogeneous variance dictionary. Each network is trained across $4$ random seeds, with the mean value shown as a solid line and the contour representing the standard deviation.
\textbf{(b)} Same as (a), for the homogeneous variance dictionary.
\textbf{(c)} Same as (a), for the heterogeneous variance dictionary, post-learning.
}
\label{fig:dl}
\end{figure}

In addition to the previously described trade-off between performance and sparsity (Figure~\ref{fig:dicos}), the robustness of the representations can be further evaluated by modifying elements in the typical activation patterns. This then allows pruning less activated coefficients to further increase sparseness, testing the code's resilience to the adversarial degradation.
We pruned coefficients with absolute values below a specific threshold, iterating from $0.001$ to $0.5$ in $6$ steps. This pruning led to a construction-induced increase in sparseness, that correlated non-linearly with a decrease in PSNR for all dictionaries, while maintaining interpretable representations (Figure~\ref{fig:cutoff}), 
The pre-learning heterogeneous variance dictionary's PSNR demonstrated significantly greater resilience to coefficient degradation than the pre-learning homogeneous variance dictionary ($p<0.05$ for pruning cutoff $c>0.3$). Post-learning, both the homogeneous and heterogeneous variance dictionaries exhibited similar PSNR, reflective of their PSNR similarities before pruning (Figure~\ref{fig:dicos}). This emphasizes the advantage of heterogeneous variance in a dictionary, whether by construction or through learning, in bolstering resilience and efficiency for encoding natural images.

Overall, these findings show that sparse codes for natural images possess highly desirable properties when incorporating heterogeneous basis functions into a sparse model: enhanced sparseness (Figure~\ref{fig:dicos}d), more evenly distributed activation (Figure~\ref{fig:learning}b), and increased resilience to code degradation (Figure~\ref{fig:cutoff}a). Yet, the differences in PSNR may not necessarily translate to perceptible differences in image quality, depending on the context and application~\cite{mozhaeva2021full}.
As such, it is necessary to investigate the potential of employing such codes in objective visual processing problems, for example, in image classification.

As a coarse analogy to a neuromorphic hierarchical sparse construction of visual processing~\cite{serre2007feedforward,schrimpf2020brain, boutin2022}, we trained a deep convolutional neural network to classify the sparse codes of natural images. The CIFAR-10 dataset, which was converted to grayscale in order to match the dimensionality of the dictionaries previously described, was sparse-coded and then classified using the Resnet-18 network, reaching a maximum top-1 accuracy of $79.20\%$ in $100$ epochs (Figure~\ref{fig:dl}, Table~\ref{table:vgg}).
After sparse coding of the dataset, but without pruning of the coefficients, a learned dictionary initialized with a heterogeneous orientation variance basis achieved the highest classification accuracy ($79.20\%$). This was followed by the pre-learned version of the network ($75.08\%$), and was higher than homogeneous variance methods. Following degradation of the sparse code ($c = 0.5$), the post-learned heterogeneous variance kept similarly high performance, unlike all the other encoding scheme which showed loss of performance. The discrepancy between the deep learning performance and the previously noted similarities in PSNR and sparseness (Figure~\ref{fig:dicos}) underscores the significance of representing variance of low-level features in complex visual models.

\begin{table}[t!]\label{table:vgg}
\renewcommand{\arraystretch}{1.3}
\caption{Mean top-1 accuracy (in $\%$) $\pm$ standard deviation across 4 random initialization of ResNet-18 for varying sparse encoding schemes of CIFAR-10. $c=0.25$ and $c=0.5$ indicate the pruning level of the sparse coefficients, as done in Figure~\ref{fig:cutoff}.}
\begin{center}
\begin{tabular}{l|ccc}
\hline
 Encoding scheme                &  No pruning    & c=0.25     & c=0.5    \\
\hline
 Homogeneous, pre-learning      &           $70.65 \pm 0.30$          & $69.70 \pm 0.26$    & $67.83 \pm 0.47$  \\
 Homogeneous, post-learning     &           $67.31 \pm 0.20$          & $66.24 \pm 0.01$    & $67.40 \pm 0.12$  \\
 Heterogeneous, pre-learning    &           $75.08 \pm 0.10$          & $71.81 \pm 0.41$    & $65.20 \pm 0.40$  \\
 Heterogeneous, post-learning   &           $\bf79.20 \pm 0.11$          & $\bf 78.98 \pm 0.00$    & $\bf 79.26 \pm 0.02$  \\
\hline
\end{tabular}
\end{center}
\end{table}

% ----------------------------------------------------------
% Discussion
% ----------------------------------------------------------
\section*{Discussion} 
% Summary
Neural systems leverage heterogeneity for increased computational efficiency~\cite{perez2021neural, di2021optimal}. Here, we have explored the effects of such heterogeneous encoding of orientation variance by integrating it into a convolutional sparse coding dictionary. Our findings show that this outperforms conventional feature-representing dictionaries with fixed variance, both in sparsity and robustness, at the cost of reconstruction performance. However, these representations can be effectively employed in subsequent visual processing stages, where they result in significantly improved performances of deep convolutional neural networks. Overall, these results imply that incorporating variance in sparse coding dictionaries can substantially improve the encoding and processing of natural images.

% LCA ties
The connection between sparse models and neural codes, which underlies the motivation behind this approach, could be further showcased using biologically plausible algorithms, such as the Locally Competitive Algorithm (LCA)~\cite{rozell2008sparse}. Rather than enforcing sparsity through convolution as done here, this model uses a mechanism of reciprocal inhibition between each of its elements, a process that mimics particular recurrent inhibition connectivity patterns observed in the cortex~\cite{coultrip1992cortical}. This method potentially mirrors a neural adaptation of winner-takes-all algorithms, reflecting innate competition and selective activation within neural networks, and highlights the potential role of feedback loops to improve sparse coding~\cite{boutin2020feedback}. Under this analogy, LCA could reinforce the presented framework of heterogeneity by extending it from features space (i.e., receptive fields) to also include the connectivity matrix (i.e., synaptic weights).
%
%New sections on neuromorphic hardware
In terms of hardware, the use of variance weighting by such a lateral inhibition mechanism could provide dynamic computational allocation for significant, unpredictable fluctuations in the data, while reducing or bypassing routine, predictable data streams. 
This arguably reflects the response characteristics and dynamics of cortical neurons~\cite{henaff2020representation, ladret2023cortical}. Emphasizing these pronounced shifts could streamline the data transmitted across physical channels, addressing a primary source of thermal and computational efficiency bottlenecks in neuromorphic hardware~\cite{eshraghian2022memristor, rahimi2020complementary}.

% Deep learning
In the context of image classification, our approach employing sparse coding achieved a top-1 accuracy of 79.20\% on the CIFAR-10 dataset. While this falls short of the state-of-the-art performance exceeding 99.0\% accuracy using color images and transformer architectures~\cite{dosovitskiy2020image}, it is important to note that our primary objective centered on comparing model performance with heterogeneous degree of variance in the initial layer, rather than solely pursuing state-of-the-art results. 
Here, the high dimensionality of the sparse-coded CIFAR-10 dataset (144 input dimensions or sparse channels), in contrast to the standard 3 dimensions in RGB images, likely contributes to this difference of accuracy. 
Direct integration of sparse coding with deep neural networks is a promising avenue of research that aligns with recent developments in the fields of unsupervised learning, object recognition, and face recognition. Some approaches have emphasized the ability of sparse coding to generate succinct, high-level representations of inputs, especially when applied as a pre-processing step for unsupervised learning with unlabeled data using L1-regularized optimization algorithms~\cite{vidya2014sparse}.
In several instances, the mechanism of sparse coding has been seamlessly integrated into deep networks. For instance, the Deep Sparse Coding framework~\cite{he2014unsupervised} maintains spatial continuity between adjacent image patches, boosting performance in object recognition. Likewise, a face recognition technique combining sparse coding neural networks with softmax classifiers effectively addresses aleatoric uncertainties, including changes in lighting, expression, posture, and low-resolution scenarios~\cite{zhang2015deep}. Classifiers relying on sparse codes, produced by lateral inhibition in an LCA, exhibit strong resistance to adversarial attacks~\cite{paiton2020selectivity}. This resilience, potentially enhanced by heterogeneous dictionaries as explored here, offers a promising avenue for research in safety-critical applications.

% Bayesian processing ties 
The empirical evidence presented here can be interpreted as an implicit Bayesian process, wherein initial beliefs about the coefficients are updated using input images to learn the variance of visual features to represent optimally (sparse) orientations. Models with explicit integration of both model and input variance have distinct advantages in that sense. Namely, this allows to maximize model performance and minimizing decision uncertainty. In contrast, we here focused on an implicit understanding of this relationship, demonstrating through a simple approach that vision models can benefit from factoring-in feature variance without explicit learning rules.
%For a complete reconfiguration, one could explore Bayesian Deep Learning~\cite{kendall2017uncertainties}, which furnishes an explicit framework but necessitates a full redesign of the existing model.

\section{Acknowledgments}
This work was supported by ANR project “AgileNeuRobot ANR-20-CE23-0021” to L.U.P, a CIHR grant to C.C (PJT-148959) and a PhD grant from \'Ecole Doctorale 62 to H.J.L.
H.J.L. would like to thank the 2023 Telluride Neuromorphic Cognition Engineering Workshop for fostering productive discussions on natural images, and for the opportunity to gather some that were used in the present research.

% ----------------------------------------------------------
% ----------------------------------------------------------
% ----------------------------------------------------------

\printbibliography 

@String(ICASSP=	{ICASSP})

@article{fischer2006sparse,
	title = {Sparse {Approximation} of {Images} {Inspired} from the {Functional} {Architecture} of the {Primary} {Visual} {Areas}},
	volume = {2007},
	copyright = {All rights reserved},
	issn = {16876172},
	url = {http://dx.doi.org/10.1155/2007/90727},
	doi = {10.1155/2007/90727},
	number = {1},
	journal = {EURASIP Journal on Advances in Signal Processing},
	author = {Fischer, Sylvain and Redondo, Rafael and Perrinet, Laurent and Cristóbal, Gabriel},
	month = dec,
	year = {2006},
	pages = {1--17},
}

@article{fair2019neuromorphic,
	title = {Sparse {Coding} {Using} the {Locally} {Competitive} {Algorithm} on the {TrueNorth} {Neurosynaptic} {System}},
	volume = {13},
	issn = {1662-453X},
	url = {https://www.frontiersin.org/articles/10.3389/fnins.2019.00754},
	urldate = {2023-12-21},
	journal = {Frontiers in Neuroscience},
	author = {Fair, Kaitlin L. and Mendat, Daniel R. and Andreou, Andreas G. and Rozell, Christopher J. and Romberg, Justin and Anderson, David V.},
	year = {2019},
}

@article{rahimi2020complementary,
  title={Complementary metal-oxide semiconductor and memristive hardware for neuromorphic computing},
  author={Rahimi Azghadi, Mostafa and Chen, Ying-Chen and Eshraghian, Jason K and Chen, Jia and Lin, Chih-Yang and Amirsoleimani, Amirali and Mehonic, Adnan and Kenyon, Anthony J and Fowler, Burt and Lee, Jack C and others},
  journal={Advanced Intelligent Systems},
  volume={2},
  number={5},
  pages={1900189},
  year={2020},
  publisher={Wiley Online Library}
}

@article{eshraghian2022memristor,
  title={Memristor-based binarized spiking neural networks: Challenges and applications},
  author={Eshraghian, Jason K and Wang, Xinxin and Lu, Wei D},
  journal={IEEE Nanotechnology Magazine},
  volume={16},
  number={2},
  pages={14--23},
  year={2022},
  publisher={IEEE}
}

@article{friston2005theory,
  title={A theory of cortical responses},
  author={Friston, Karl},
  journal={Philosophical transactions of the Royal Society B: Biological sciences},
  volume={360},
  number={1456},
  pages={815--836},
  year={2005},
  publisher={The Royal Society London}
}

@article{laughlin1981simple,
  title={A simple coding procedure enhances a neuron's information capacity},
  author={Laughlin, Simon},
  journal={Zeitschrift f{\"u}r Naturforschung c},
  volume={36},
  number={9-10},
  pages={910--912},
  year={1981},
  publisher={Verlag der Zeitschrift f{\"u}r Naturforschung}
}

@article{stringer2019high,
  title={High-dimensional geometry of population responses in visual cortex},
  author={Stringer, Carsen and Pachitariu, Marius and Steinmetz, Nicholas and Carandini, Matteo and Harris, Kenneth D},
  journal={Nature},
  volume={571},
  number={7765},
  pages={361--365},
  year={2019},
  publisher={Nature Publishing Group UK London}
}

@article{dosovitskiy2020image,
  title={An image is worth $16\times 16$ words: Transformers for image recognition at scale},
  author={Dosovitskiy, Alexey and Beyer, Lucas and Kolesnikov, Alexander and Weissenborn, Dirk and Zhai, Xiaohua and Unterthiner, Thomas and Dehghani, Mostafa and Minderer, Matthias and Heigold, Georg and Gelly, Sylvain and others},
  journal={arXiv preprint arXiv:2010.11929},
  year={2020}
}

@article{olshausen1996emergence,
  title={Emergence of simple-cell receptive field properties by learning a sparse code for natural images},
  author={Olshausen, Bruno A and Field, David J},
  journal={Nature},
  volume={381},
  number={6583},
  pages={607--609},
  year={1996},
  publisher={Nature Publishing Group}
}

@article{olshausen1997sparse,
  title={Sparse coding with an overcomplete basis set: A strategy employed by V1?},
  author={Olshausen, Bruno A and Field, David J},
  journal={Vision research},
  volume={37},
  number={23},
  pages={3311--3325},
  year={1997},
  publisher={Elsevier}
}

@article{wohlberg2015efficient,
  title={Efficient algorithms for convolutional sparse representations},
  author={Wohlberg, Brendt},
  journal={IEEE Transactions on Image Processing},
  volume={25},
  number={1},
  pages={301--315},
  year={2015},
  publisher={IEEE}
}

@article{lee2006efficient,
  title={Efficient sparse coding algorithms},
  author={Lee, Honglak and Battle, Alexis and Raina, Rajat and Ng, Andrew},
  journal={Advances in neural information processing systems},
  volume={19},
  year={2006}
}

@inproceedings{wohlberg2014efficient,
  title={Efficient convolutional sparse coding},
  author={Wohlberg, Brendt},
  booktitle={2014 IEEE International Conference on Acoustics, Speech and Signal Processing (ICASSP)},
  pages={7173--7177},
  year={2014},
  organization={IEEE}
}

@article{chen2001atomic,
  title={Atomic decomposition by basis pursuit},
  author={Chen, Scott Shaobing and Donoho, David L and Saunders, Michael A},
  journal={SIAM review},
  volume={43},
  number={1},
  pages={129--159},
  year={2001},
  publisher={SIAM}
}

@article{serre2007feedforward,
  title={A feedforward architecture accounts for rapid categorization},
  author={Serre, Thomas and Oliva, Aude and Poggio, Tomaso},
  journal={Proceedings of the national academy of sciences},
  volume={104},
  number={15},
  pages={6424--6429},
  year={2007},
  publisher={National Acad Sciences}
}

@article{goris2015origin,
  title={Origin and function of tuning diversity in macaque visual cortex},
  author={Goris, Robbe LT and Simoncelli, Eero P and Movshon, J Anthony},
  journal={Neuron},
  volume={88},
  number={4},
  pages={819--831},
  year={2015},
  publisher={Elsevier}
}

@article{perez2021neural,
  title={Neural heterogeneity promotes robust learning},
  author={Perez-Nieves, Nicolas and Leung, Vincent CH and Dragotti, Pier Luigi and Goodman, Dan FM},
  journal={Nature communications},
  volume={12},
  number={1},
  pages={5791},
  year={2021},
  publisher={Nature Publishing Group UK London}
}

@article{di2021optimal,
  title={Optimal responsiveness and information flow in networks of heterogeneous neurons},
  author={Di Volo, Matteo and Destexhe, Alain},
  journal={Scientific reports},
  volume={11},
  number={1},
  pages={17611},
  year={2021},
  publisher={Nature Publishing Group UK London}
}

@article{paiton2020selectivity,
  title={Selectivity and robustness of sparse coding networks},
  author={Paiton, Dylan M and Frye, Charles G and Lundquist, Sheng Y and Bowen, Joel D and Zarcone, Ryan and Olshausen, Bruno A},
  journal={Journal of vision},
  volume={20},
  number={12},
  pages={10--10},
  year={2020},
  publisher={The Association for Research in Vision and Ophthalmology}
}

@article{fischer2007self,
  title={Self-invertible 2D log-Gabor wavelets},
  author={Fischer, Sylvain and {\v{S}}roubek, Filip and Perrinet, Laurent U and Redondo, Rafael and Crist{\'o}bal, Gabriel},
  journal={International Journal of Computer Vision},
  volume={75},
  number={2},
  pages={231--246},
  year={2007},
  publisher={Springer}
}

@article{fischer2007sparse,
	author = {Fischer, Sylvain and Redondo, Rafael and Perrinet, Laurent and Crist{\'o}bal, Gabriel and Cristobal, Gabriel and Crist{\'o}bal, Gabriel and Cristobal, Gabriel},
	year = {2007},
	ids = {Fischer07,Fischer07a},
	issn = {16876172},
	journal = {EURASIP Journal on Advances in Signal Processing},
	number = {1},
	pages = {1--17},
	publisher = {{Hindawi Publishing Corp.}},
	title = {Sparse {{Approximation}} of {{Images Inspired}} from the {{Functional Architecture}} of the {{Primary Visual Areas}}},
	volume = {2007},}

@article{lewicki1998coding,
  title={Coding time-varying signals using sparse, shift-invariant representations},
  author={Lewicki, Michael and Sejnowski, Terrence J},
  journal={Advances in neural information processing systems},
  volume={11},
  year={1998}
}

@article{van2014scikit,
  title={scikit-image: image processing in Python},
  author={Van der Walt, Stefan and Sch{\"o}nberger, Johannes L and Nunez-Iglesias, Juan and Boulogne, Fran{\c{c}}ois and Warner, Joshua D and Yager, Neil and Gouillart, Emmanuelle and Yu, Tony},
  journal={PeerJ},
  volume={2},
  pages={e453},
  year={2014},
  publisher={PeerJ Inc.}
}

@inproceedings{he2016deep,
  title={Deep residual learning for image recognition},
  author={He, Kaiming and Zhang, Xiangyu and Ren, Shaoqing and Sun, Jian},
  booktitle={Proceedings of the IEEE conference on computer vision and pattern recognition},
  pages={770--778},
  year={2016}
}

@article{schrimpf2020brain,
  title={Brain-score: Which artificial neural network for object recognition is most brain-like?},
  author={Schrimpf, Martin and Kubilius, Jonas and Hong, Ha and Majaj, Najib J and Rajalingham, Rishi and Issa, Elias B and Kar, Kohitij and Bashivan, Pouya and Prescott-Roy, Jonathan and Geiger, Franziska and others},
  journal={BioRxiv},
  pages={407007},
  year={2020},
  publisher={Cold Spring Harbor Laboratory}
}

@article{henaff2020representation,
  title={Representation of visual uncertainty through neural gain variability},
  author={H{\'e}naff, Olivier J and Boundy-Singer, Zoe M and Meding, Kristof and Ziemba, Corey M and Goris, Robbe LT},
  journal={Nature communications},
  volume={11},
  number={1},
  pages={1--12},
  year={2020},
  publisher={Nature Publishing Group}
}

@article{simoncelli2001natural,
  title={Natural image statistics and neural representation},
  author={Simoncelli, Eero P and Olshausen, Bruno A},
  journal={Annual review of neuroscience},
  volume={24},
  number={1},
  pages={1193--1216},
  year={2001},
  publisher={Annual Reviews 4139 El Camino Way, PO Box 10139, Palo Alto, CA 94303-0139, USA}
}

@article{field1987relations,
  title={Relations between the statistics of natural images and the response properties of cortical cells},
  author={Field, David J},
  journal={Josa a},
  volume={4},
  number={12},
  pages={2379--2394},
  year={1987},
  publisher={Optical Society of America}
}

@article{swindale1998orientation,
  title={Orientation tuning curves: empirical description and estimation of parameters},
  author={Swindale, Nicholas V},
  journal={Biological cybernetics},
  volume={78},
  number={1},
  pages={45--56},
  year={1998},
  publisher={Springer}
}

@article{rozell2008sparse,
  title={Sparse coding via thresholding and local competition in neural circuits},
  author={Rozell, Christopher J and Johnson, Don H and Baraniuk, Richard G and Olshausen, Bruno A},
  journal={Neural computation},
  volume={20},
  number={10},
  pages={2526--2563},
  year={2008},
  publisher={MIT Press One Rogers Street, Cambridge, MA 02142-1209, USA journals-info~…}
}

@article{coultrip1992cortical,
  title={A cortical model of winner-take-all competition via lateral inhibition},
  author={Coultrip, Robert and Granger, Richard and Lynch, Gary},
  journal={Neural networks},
  volume={5},
  number={1},
  pages={47--54},
  year={1992},
  publisher={Elsevier}
}

@article{ladret2023cortical,
  title={Cortical recurrence supports resilience to sensory variance in the primary visual cortex},
  author={Ladret, Hugo J and Cortes, Nelson and Ikan, Lamyae and Chavane, Fr{\'e}d{\'e}ric and Casanova, Christian and Perrinet, Laurent U},
  journal={Communications Biology},
  volume={6},
  number={1},
  pages={667},
  year={2023},
  publisher={Nature Publishing Group UK London}
}

@incollection{perrinet2015sparse,
	author = {Perrinet, Laurent U},
	booktitle = {Biologically {{Inspired Computer Vision}}},
	year = {2015},
	editor = {Keil, Matthias and Crist{\'o}bal, Gabriel and Perrinet, Laurent U},
	ids = {Perrinet15bicv,Perrinet2015c},
	location = {{Weinheim, Germany}},
	pages = {319--346},
	publisher = {{Wiley-VCH Verlag GmbH \& Co. KGaA}},
	title = {Sparse {{Models}} for {{Computer Vision}}},
}

@article{pettypiece2010integration,
  title={Integration of haptic and visual size cues in perception and action revealed through cross-modal conflict},
  author={Pettypiece, Charles E and Goodale, Melvyn A and Culham, Jody C},
  journal={Experimental brain research},
  volume={201},
  number={4},
  pages={863--873},
  year={2010},
  publisher={Springer}
}

@inproceedings{zhang2015deep,
  title={Deep neural network for face recognition based on sparse autoencoder},
  author={Zhang, Zhuomin and Li, Jing and Zhu, Renbing},
  booktitle={2015 8th International Congress on Image and Signal Processing (CISP)},
  pages={594--598},
  year={2015},
  organization={IEEE}
}

@inproceedings{he2014unsupervised,
  title={Unsupervised feature learning by deep sparse coding},
  author={He, Yunlong and Kavukcuoglu, Koray and Wang, Yun and Szlam, Arthur and Qi, Yanjun},
  booktitle={Proceedings of the 2014 SIAM international conference on data mining},
  pages={902--910},
  year={2014},
  organization={SIAM}
}

@inproceedings{vidya2014sparse,
  title={Sparse coding: a deep learning using unlabeled data for high-level representation},
  author={Vidya, Raghavendran and Nasira, GM and Priyankka, RP Jaia},
  booktitle={2014 World Congress on Computing and Communication Technologies},
  pages={124--127},
  year={2014},
  organization={IEEE}
}

@article{gousseau2001natural,
  title={Are natural images of bounded variation?},
  author={Gousseau, Yann and Morel, Jean-Michel},
  journal={SIAM Journal on Mathematical Analysis},
  volume={33},
  number={3},
  pages={634--648},
  year={2001},
  publisher={SIAM}
}

@inproceedings{nakamura2015robot,
  title={Robot audition based acoustic event identification using a bayesian model considering spectral and temporal uncertainties},
  author={Nakamura, Keisuke and Nakadai, Kazuhiro},
  booktitle={2015 IEEE/RSJ International Conference on Intelligent Robots and Systems (IROS)},
  pages={4840--4845},
  year={2015},
  organization={IEEE}
}

@article{appelle1972perception,
  title={Perception and discrimination as a function of stimulus orientation: the" oblique effect" in man and animals.},
  author={Appelle, Stuart},
  journal={Psychological bulletin},
  volume={78},
  number={4},
  pages={266},
  year={1972},
  publisher={American Psychological Association}
}

@article{orban2016neural,
  title={Neural variability and sampling-based probabilistic representations in the visual cortex},
  author={Orb{\'a}n, Gerg{\H{o}} and Berkes, Pietro and Fiser, J{\'o}zsef and Lengyel, M{\'a}t{\'e}},
  journal={Neuron},
  volume={92},
  number={2},
  pages={530--543},
  year={2016},
  publisher={Elsevier}
}

@article{wang2019global,
  title={Global convergence of ADMM in nonconvex nonsmooth optimization},
  author={Wang, Yu and Yin, Wotao and Zeng, Jinshan},
  journal={Journal of Scientific Computing},
  volume={78},
  number={1},
  pages={29--63},
  year={2019},
  publisher={Springer}
}

@inproceedings{mozhaeva2021full,
  title={Full reference video quality assessment metric on base human visual system consistent with PSNR},
  author={Mozhaeva, Anastasia and Streeter, Lee and Vlasuyk, Igor and Potashnikov, Aleksei},
  booktitle={2021 28th Conference of Open Innovations Association (FRUCT)},
  pages={309--315},
  year={2021},
  organization={IEEE}
}

@article{hubel1962receptive,
  title={Receptive fields, binocular interaction and functional architecture in the cat's visual cortex},
  author={Hubel, David H and Wiesel, Torsten N},
  journal={The Journal of physiology},
  volume={160},
  number={1},
  pages={106},
  year={1962},
  publisher={Wiley-Blackwell}
}

@article{kingma2014adam,
  title={Adam: A method for stochastic optimization},
  author={Kingma, Diederik P and Ba, Jimmy},
  journal={arXiv preprint arXiv:1412.6980},
  year={2014}
}

@article{hansen2004horizontal,
  title={A horizontal bias in human visual processing of orientation and its correspondence to the structural components of natural scenes},
  author={Hansen, Bruce C and Essock, Edward A},
  journal={Journal of vision},
  volume={4},
  number={12},
  pages={5--5},
  year={2004},
  publisher={The Association for Research in Vision and Ophthalmology}
}

@inproceedings{wohlberg2017sporco,
  title={SPORCO: A Python package for standard and convolutional sparse representations},
  author={Wohlberg, Brendt},
  booktitle={Proceedings of the 15th Python in Science Conference, Austin, TX, USA},
  pages={1--8},
  year={2017}
}

@article{ladret2023imgs,
    author = {Ladret, Hugo},
    title = {HD natural images database for sparse coding},
    year = {2023},
    journal = {FigShare},
    doi = {"10.6084/m9.figshare.24167265.v1"}
}

@article{coppola1998distribution,
  title={The distribution of oriented contours in the real world},
  author={Coppola, David M and Purves, Harriett R and McCoy, Allison N and Purves, Dale},
  journal={Proceedings of the National Academy of Sciences},
  volume={95},
  number={7},
  pages={4002--4006},
  year={1998},
  publisher={National Acad Sciences}
}

@book{helmholtz1867treatise,
  title={Treatise on physiological optics},
  author={Helmholtz, Hermann LF von},
  year={1867},
  publisher={}
}

@article{ruderman1994statistics,
  title={The statistics of natural images},
  author={Ruderman, Daniel L},
  journal={Network: computation in neural systems},
  volume={5},
  number={4},
  pages={517},
  year={1994},
  publisher={IOP Publishing}
}

@article{hullermeier2021aleatoric,
  title={Aleatoric and epistemic uncertainty in machine learning: An introduction to concepts and methods},
  author={H{\"u}llermeier, Eyke and Waegeman, Willem},
  journal={Machine Learning},
  volume={110},
  number={3},
  pages={457--506},
  year={2021},
  publisher={Springer}
}

@article{boutin2020feedback,
	author = {Boutin, Victor and Franciosini, Angelo and Ruffier, Franck and Perrinet, Laurent U},
	ids = {BoutinFranciosiniRuffierPerrinet20},
	journal = {Neural Computation},
	number = {11},
	pages = {2279--2309},
	publisher = {{MIT Press}},
	title = {Effect of Top-down Connections in {{Hierarchical Sparse Coding}}},
	volume = {32},
	year = {2020-02-04, November 2020},}

@article{boutin2020sparse,
	author = {Boutin, Victor and Franciosini, Angelo and Chavane, Fr{\'e}d{\'e}ric Y and Ruffier, Franck and Perrinet, Laurent U},
	grants = {doc-2-amu,phd-icn,mesocentre},
	journal = {PLoS Computational Biology},
	publisher = {{Public Library of Science San Francisco, CA USA}},
	title = {Sparse {{Deep Predictive Coding}} Captures Contour Integration Capabilities of the Early Visual System},
	date = {2020-05-12},
	year = {2020}}

@article{boutin2022,
	author = {Boutin, Victor and Franciosini, Angelo and Chavane, Fr{\'e}d{\'e}ric and Perrinet, Laurent U.},
	doi = {10.1371/journal.pcbi.1010270},
	issn = {1553-7358},
	journal = {PLOS Computational Biology},
	number = {7},
	pages = {e1010270},
	publisher = {{Public Library of Science}},
	shortjournal = {PLOS Computational Biology},
	title = {Pooling Strategies in {{V1}} Can Account for the Functional and Structural Diversity across Species},
	url = {https://journals.plos.org/ploscompbiol/article?id=10.1371/journal.pcbi.1010270},
	urldate = {2022-09-14},
	volume = {18},
	year = {2022}}

% ----------------------------------------------------------
% ----------------------------------------------------------
% ----------------------------------------------------------

\newpage

\renewcommand{\figurename}{Appendix A Figure}
\setcounter{figure}{0} 
\section*{Appendix A - Additional Convolutional Sparse Coding details}\label{apdxA}
\begin{figure}[ht]
    \centering
    \includegraphics[scale = 0.45]{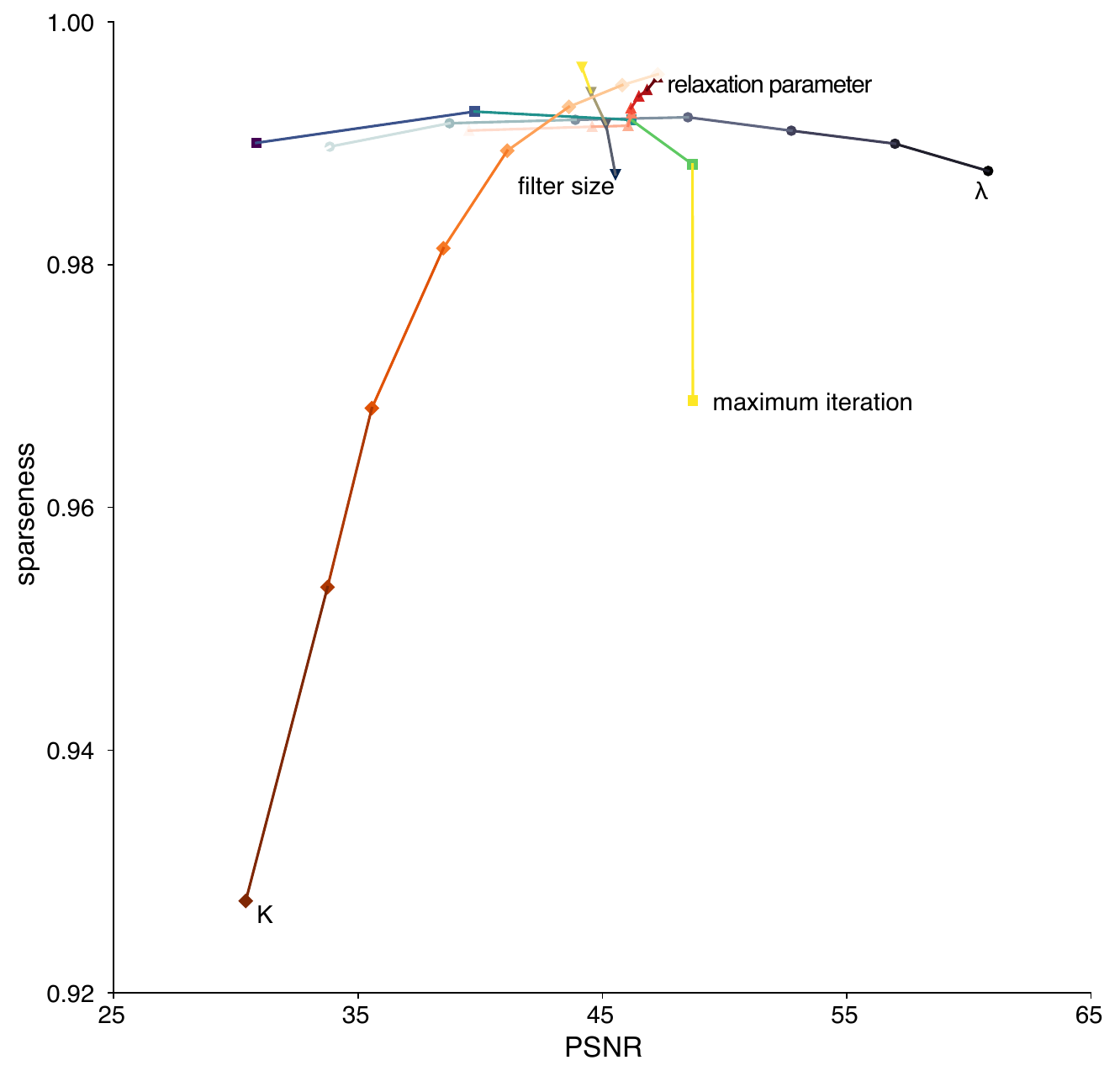}
    \caption{
    Parametrization of the CSC learning algorithm.
    $\lambda$ was varied in 8 steps in a $[0.001:0.1]$ range,
    max iteration in 5 steps in a $[10:1000]$ range, 
    relaxation parameter $\rho$ in 8 steps in a $[0.2:1.8]$ range,
    filter size in 8 steps in a $[5:21]$ pixels range and
    $K$ in 8 steps in a $[89:2351]$ range.
    }
    \label{fig:param_scan}
    \end{figure}
    
Convolutional Sparse Coding was implemented using an Alternating Direction Method of Multipliers (ADMM) algorithm, which decomposes the problem into a standard form:
\begin{equation}
    \underset{x,y}{\operatorname{arg min}}
    f(x) + g(y)
\end{equation}
with the constraint $x = y$. This is then solved iteratively by alternating between the two sub-problems:
\begin{equation} \label{eq:admm_1}
    x_{i+1} = \underset{x}{\operatorname{arg min}} f(x)
    + \frac{\rho}{2} || x + y_i + \text{u}_i||^2_2
\end{equation}
\begin{equation} \label{eq:admm_2}
    y_{i+1} = \underset{y}{\operatorname{arg min}} g(y)
    + \frac{\rho}{2} || x_{i+1} + y + \text{u}_i||^2_2
\end{equation}
where $\rho$ is a penalty parameter that controls the convergence rate of the iterations, also called the relaxation parameter. $x$ and $y$ are residuals whose equality is enforced by the prediction error:
\begin{equation} \label{eq:admm_3}
    \text{u}_{i+1} = \text{u}_i + x_{i+1} + y_{i+1}
\end{equation}
ADMM can be readily applied to equation (\ref{eq:csc}) by introducing an auxiliary variable $Y$~\cite{wohlberg2014efficient}, such that the problem to solve becomes:
\begin{equation} \label{eq:csc_admm}
    \underset{\{x_k\},\{y_k\}}{\operatorname{arg min}} \frac{1}{2}
    || \sum_{k=1}^K \text{d}_k \ast x_k - s ||^2_2 + 
    \lambda \sum_{k=1}^K ||y_k||_1
    \text{ s.t.\ x$_k=$y$_k$}
\end{equation}
which, following the ADMM alternation in equations (\ref{eq:admm_1})-(\ref{eq:admm_3}), is solved by alternating:
\begin{equation}
    \{x_k\}_{i+1} = \underset{\{x_k\}}{\operatorname{arg min}} 
    \frac{1}{2} || \sum_{k=1}^K \text{d}_k * x_k - s ||^2_2
    + \frac{\rho}{2} || x_k - y_{k,i} + \text{u}_{k,i}||^2_2
\end{equation}
\begin{equation}
    \{y_k\}_{i+1} = \underset{\{y_k\}}{\operatorname{arg min}} 
    \lambda \sum_{k=1}^K || y_k ||_1
    + \frac{\rho}{2} || x_{k,i+1} - y_{k} + \text{u}_{k,i}||^2_2
\end{equation}
\begin{equation}
    \text{u}_{k,i+1} = \text{u}_{k,i} + x_{k,i+1} - y_{k,i+1}
\end{equation}

% ----------------------------------------------------------
% ----------------------------------------------------------
% ----------------------------------------------------------

\newpage
\renewcommand{\figurename}{Appendix B Figure}
\setcounter{figure}{0} 
\section*{Appendix B - Homogeneous variance dictionary} \label{apdxC}
Results from the main text are shown here for the homogeneous variance dictionary, post-learning. 
\begin{figure}[ht]
\centering
\includegraphics[width=\linewidth]{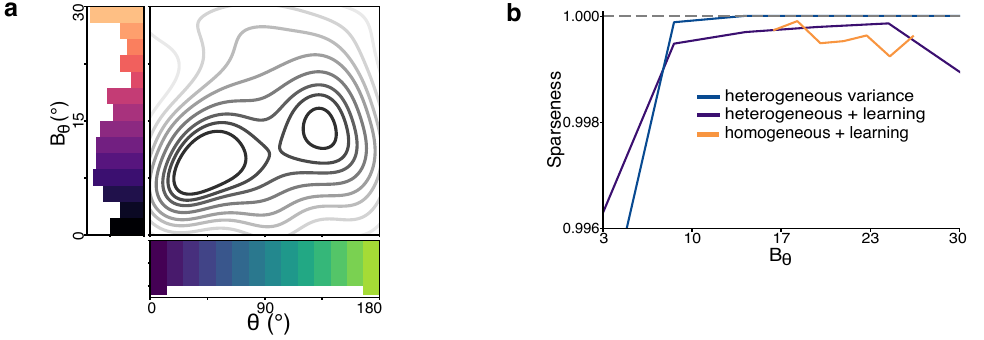}
\caption{
Learning balances coefficient distribution.
\textbf{(a)} Kernel density estimation of coefficients over $\theta_0$ and $B_\theta$ after learning from the homogeneous variance dictionary.
\textbf{(b)} Sparseness of coefficients for each $B_\theta$. Sparseness $=1$ is represented as a gray dashed line.
}
\label{supfig:learning}
\end{figure}

\begin{figure}[h]
\centering
\includegraphics[width=\linewidth]{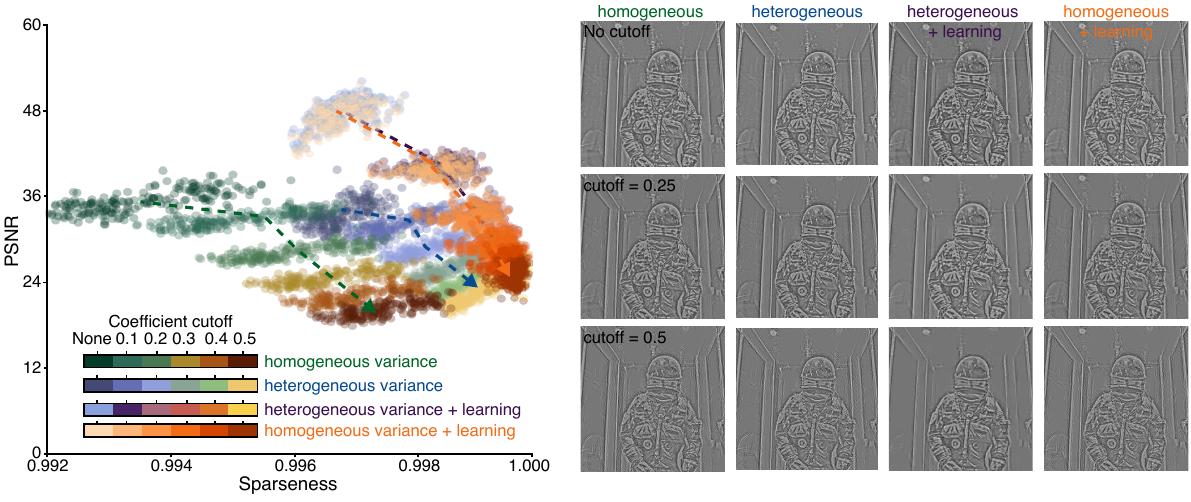}
\caption{
Sparse coefficients can be pruned to boost sparsity.
\textbf{(a)} Pruning of the coefficients based on their values and resulting sparseness/PSNR for both dictionaries.
\textbf{(b)} Reconstruction of the image shown in Figure~\ref{fig:intro} with different cutoff levels.
}
\label{supfig:cutoff}
\end{figure}

\end{document}